\documentclass[pra,aps,showpacs,floatfix,twocolumn,nofootinbib,superscriptaddress]{revtex4-1}
\usepackage{amsmath}
\usepackage{amsfonts}
\usepackage{amssymb}
\usepackage{revsymb}
\usepackage{graphicx}
\allowdisplaybreaks

\def\ket#1{|\,#1 \,\rangle}
\def\bra#1{\langle \, #1 \,|}

\begin{document}
\title{Oscillating quadrupole effects in high precision metrology.}
\author{K. J. Arnold}
\affiliation{Centre for Quantum Technologies, National University of Singapore, 3 Science Drive 2, 117543 Singapore}
\author{R. Kaewuan}
\affiliation{Centre for Quantum Technologies, National University of Singapore, 3 Science Drive 2, 117543 Singapore}
\author{T. R. Tan}
\affiliation{Centre for Quantum Technologies, National University of Singapore, 3 Science Drive 2, 117543 Singapore}
\affiliation{Department of Physics, National University of Singapore, 2 Science Drive 3, 117551 Singapore}
\author{M. D. Barrett}
\affiliation{Centre for Quantum Technologies, National University of Singapore, 3 Science Drive 2, 117543 Singapore}
\affiliation{Department of Physics, National University of Singapore, 2 Science Drive 3, 117551 Singapore}
\email{phybmd@nus.edu.sg}
\begin{abstract}
The influence of oscillating quadrupole fields on atomic energy levels is examined theoretically and general expressions for the quadrupole matrix elements are given.  The results are relevant to any ion-based clock in which one of the clock states supports a quadrupole moment.    Clock shifts are estimated for $^{176}$Lu$^+$ and indicate that coupling to the quadrupole field would not be a limitation to clock accuracy at the $\lesssim10^{-19}$ level.  Nevertheless, a method is suggested that would allow this shift to be calibrated.  This method utilises a resonant quadrupole coupling that enables the quadrupole moment of the atom to be measured.  A proof-of-principle demonstration is given using $^{138}$Ba$^+$, in which the quadrupole moment of the $D_{5/2}$ state is estimated to be $\Theta=3.229(89) e a_0^2$.
\end{abstract}
\maketitle
In a recent paper \cite{gan2018oscillating}, the effects of oscillating magnetic fields in high precision metrology were explored.  In that work it was shown that magnetic fields driven by the oscillating potential of a Paul trap could have a significant influence on high precision measurements and optical atomic clocks.  Given that the oscillating potential itself provides a strong quadrupole field, it is of interest to consider the effects this might have on energy levels supporting a non-zero quadrupole moment.

The interaction of external electric-field gradients with the quadrupole moment of the atom is described by tensor operators of rank two \cite{ItanoQuad}.  As for ac magnetic fields, the interaction couples levels primarily within the same fine-structure manifold.  However, the rank-2 operators provide a coupling between levels having $\Delta F=0,\pm1,\pm2$ and $\Delta m=0,\pm1,\pm2$.  In this paper, a general expression for the interaction matrix elements is derived and the various level shifts and effects that can occur are considered.  These results can be readily applied to any system.  For the purposes of illustration, fractional frequency shifts for three clock transitions of $^{176}$Lu$^+$ are estimated for experimentally relevant parameter values.
\section{Theory}
The notations and conventions used here follow that used in \cite{ItanoQuad}.  The principal-axis (primed) frame $(x',y',z')$ is one in which the electric potential in the neighbourhood of the atom has the simple form
\begin{equation}
\label{principalEP}
\Phi(x',y',z')=A(x'^2+y'^2-2z'^2)+\epsilon(x'^2-y'^2),
\end{equation}
while a laboratory (unprimed) frame $(x,y,z)$ is one in which the magnetic field is oriented along the $z$ axis.  Using this form of the potential,
the time-dependent potential associated with an ideal linear Paul trap has $A=0$ and that for the ideal quadrupole trap has $\epsilon=0$, with the time-dependence provided by a $\cos(\Omega_\mathrm{rf} t)$ factor, where $\Omega_\mathrm{rf}$ is the trap drive frequency.

In the principal-axis frame, the spherical components of $\nabla\mathbf{E}^{(2)}$ are
\begin{equation}
\nabla E_0^{(2)'}=-2A, \quad \nabla E_{\pm1}^{(2)'}=0, \quad \nabla E_{\pm2}^{(2)'}=\epsilon \sqrt{\frac{2}{3}}.
\end{equation}
and the interaction $H_Q$ has the simple form
\begin{equation}
\label{quadInteraction}
H_Q=-2 A \Theta_0^{(2)'}+\epsilon \sqrt{\frac{2}{3}}\left(\Theta_{2}^{(2)'}+\Theta_{-2}^{(2)'}\right).
\end{equation}
States $\ket{JFm}'$ defined in the principal-axis frame and states $\ket{JF\mu}$ defined in the laboratory frame are related by
\begin{equation}
\ket{JFm}'=\sum_\mu D_{\mu, m}^{(F)}(\boldsymbol{\omega})\ket{JF\mu},
\end{equation}
with the inverse relation
\begin{equation}
\ket{JF\mu}=\sum_m D_{\mu, m}^{(F)^*}(\boldsymbol{\omega})\ket{JFm}',
\end{equation}
where $\boldsymbol{\omega}$ denotes a set of Euler angles $\{\alpha,\beta,\gamma\}$ taking the principal-axis frame to the laboratory frame defined with the same convention used in \cite{ItanoQuad}.  Specifically, starting from the principle axis frame, the coordinate system is rotated about $z$ by $\alpha$, then about the new $y$ axis by $\beta$ and then about the new $z$ axis by $\gamma$ so that the rotated coordinate system coincides with the laboratory coordinate system.  As the rotation $\gamma$ is parallel to the magnetic field, it has no effect and can be set to zero.  The rotation matrices $D_{\mu, m}^{(F)}(\boldsymbol{\omega})$ are given in the passive interpretation for which expressions can be found in \cite{edmonds2016angular}.

Matrix elements of $\Theta_{q}^{(2)'}$ in the laboratory frame can then found using the same derivation given in \cite{ItanoQuad}, generalized to include off-diagonal matrix elements. Explicitly
\begin{widetext}
\begin{align}
\bra{JF'\mu'}\Theta_{q}^{(2)'}\ket{JF\mu}&=\sum_{m',m} D_{\mu',m'}^{(F')}(\boldsymbol{\omega}) D_{\mu, m }^{(F)^*}(\boldsymbol{\omega})\; {^\backprime\bra{JF'm'}}\Theta_{q}^{(2)'}\ket{JFm}'\\
&=\langle{JF'}\|\Theta^{(2)}\|JF\rangle\sum_{m',m} D_{\mu',m'}^{(F')}(\boldsymbol{\omega}) D_{\mu, m }^{(F)^*}(\boldsymbol{\omega})(-1)^{F'-m'}\begin{pmatrix}F' & 2 & F\\-m' & q & m\end{pmatrix}\\
&=\langle{JF'}\|\Theta^{(2)}\|JF\rangle\sum_{m',m} D_{\mu',m'}^{(F')}(\boldsymbol{\omega}) D_{-\mu, -m }^{(F)}(\boldsymbol{\omega})(-1)^{F'-m'+\mu-m}\begin{pmatrix}F' & 2 & F\\-m' & q & m\end{pmatrix}\\
&=(-1)^{F'-\mu-q}\langle{JF'}\|\Theta^{(2)}\|JF\rangle\sum_{m',m} D_{\mu',m'}^{(F')}(\boldsymbol{\omega}) D_{-\mu, -m }^{(F)}(\boldsymbol{\omega})\begin{pmatrix}F' & 2 & F\\-m' & q & m\end{pmatrix}\\
&=(-1)^{F'-\mu-q}\langle{JF'}\|\Theta^{(2)}\|JF\rangle\nonumber\\
&\qquad\times\sum_{\substack{m',m\\K,n',n}} (2K+1) \begin{pmatrix}F' & 2 & F\\-m' & q & m\end{pmatrix}\begin{pmatrix}F' & F & K\\ \mu' & -\mu & n'\end{pmatrix}\begin{pmatrix}F' & F & K\\ m' & -m & n\end{pmatrix}D_{n',n}^{(K)^*}(\boldsymbol{\omega})\\
&=(-1)^{F'-\mu-q}\langle{JF'}\|\Theta^{(2)}\|JF\rangle\nonumber\\
& \qquad\times\sum_{\substack{m',m\\K,n',n}} (2K+1) \begin{pmatrix}F' & F & K\\ \mu' & -\mu & n'\end{pmatrix}\begin{pmatrix}F' & F & 2\\ m' & -m & -q\end{pmatrix}\begin{pmatrix}F' & F & K\\ m' & -m & n\end{pmatrix}D_{n',n}^{(K)^*}(\boldsymbol{\omega})\\
&=(-1)^{F'-\mu-q}\langle{JF'}\|\Theta^{(2)}\|JF\rangle\sum_{n'} \begin{pmatrix}F' & 2 & F\\ -\mu' & -n' & \mu \end{pmatrix} D_{n',-q}^{(2)^*}(\boldsymbol{\omega})\\
&=(-1)^{F'-\mu-q}\langle{JF'}\|\Theta^{(2)}\|JF\rangle \begin{pmatrix}F' & 2 & F\\ -\mu' & \Delta \mu & \mu \end{pmatrix} D_{-\Delta\mu,-q}^{(2)^*}(\boldsymbol{\omega})\\
&=(-1)^{F+\mu'}\langle{JF'}\|\Theta^{(2)}\|JF\rangle \begin{pmatrix}F & 2 & F'\\ \mu & \Delta \mu & -\mu' \end{pmatrix} D_{\Delta\mu,q}^{(2)}(\boldsymbol{\omega}),
\end{align}
where $\Delta\mu=\mu'-\mu$.  Using the $IJ$-coupling approximation, the reduced matrix element may be written in terms of the usual quadrupole moment $\Theta(J)=\bra{JJ}\Theta_0^{(2)}\ket{JJ}$ giving the final expression
\begin{multline}
\label{GME}
\bra{(IJ)F'\mu'}\Theta_{q}^{(2)'}\ket{(IJ)F\mu}=(-1)^{F'+F+I+J+\mu'}\sqrt{(2F'+1)(2F+1)}\\
\times \begin{Bmatrix}F & F' & 2\\J & J &I\end{Bmatrix}\begin{pmatrix}F & 2 & F'\\ \mu & \Delta \mu & -\mu' \end{pmatrix} \begin{pmatrix}J & 2 & J\\ -J & 0 & J \end{pmatrix}^{-1} \Theta(J) D_{\Delta\mu,q}^{(2)}(\boldsymbol{\omega}),
\end{multline}
where $I$ is included in the notation to identify the ordering of the $IJ$ coupling.
\end{widetext}
The only $q$-dependence for the matrix element appears in the Wigner rotation matrices and, from \cite{edmonds2016angular}
\begin{subequations}
\label{Wigner}
\begin{align}
D_{0,2}^{(2)}(\boldsymbol{\omega})+D_{0,-2}^{(2)}(\boldsymbol{\omega})&=\sqrt{\frac{3}{2}}\sin^2\beta\cos2\alpha, \\
\label{linear1}
D_{\pm1,2}^{(2)}(\boldsymbol{\omega})+D_{\pm1,-2}^{(2)}(\boldsymbol{\omega})&=\mp(\cos\beta\sin\beta \cos2\alpha\nonumber\\
&\qquad \pm i \sin\beta \sin2\alpha),\\
\label{linear2}
D_{\pm2,2}^{(2)}(\boldsymbol{\omega})+D_{\pm2,-2}^{(2)}(\boldsymbol{\omega})&=\frac{1}{2}\left(1+\cos^2\beta\right)\cos2\alpha\nonumber \\
&\qquad \pm i \cos\beta \sin2\alpha,\\
D_{0,0}^{(2)}(\boldsymbol{\omega}) &=\frac{1}{2}\left(3\cos^2\beta-1\right),\\
D_{\pm1, 0}^{(2)}(\boldsymbol{\omega}) & = \pm\sqrt{\frac{3}{2}}\;\sin\beta \cos\beta,\\
D_{\pm2, 0}^{(2)}(\boldsymbol{\omega}) & = \sqrt{\frac{3}{8}}\;\sin^2 \beta.
\end{align}
\end{subequations}

Equations ~\ref{GME} and \ref{Wigner} can then be used to determine the various influences of the time-varying trapping potential by considering the time-dependent interaction $H_Q \cos(\Omega_\mathrm{rf} t)$.  This implicitly assumes that there are no phase shifts between sources determining the electric potential given in Eq.~\ref{principalEP}.  If this were not the case, Eq.~\ref{principalEP} would have to be modified to account for a non-zero electric field at the ion position.  In the current context, such modifications would have negligible effect in any practical setup: a small phase shift between electrodes would result in a correspondingly small correction to Eq.~\ref{principalEP} and, for a trap with reasonable symmetry, correction terms would have a near-zero electric-field gradient in the neighbourhood of the ion anyway.

As with an oscillating magnetic field, the oscillating quadrupole field will (i) modulate energy levels giving rise to rf sidebands when transitions connected to the level are driven, (ii) drive resonances between levels having an energy difference matching the trap drive frequency, and (iii) shift energy levels due to off-resonant coupling to other levels.  These effects will have a complicated dependence on trap geometry and orientation with respect to the laboratory frame.  For a given setup, this can be readily calculated, but general expressions are not so illuminating.  Therefore specific examples will be used to illustrate key considerations.

To determine the typical scale of the quadrupole shift, first note that the ideal linear Paul trap ($A=0$) has
\begin{equation}
\label{epsilon}
\epsilon=\frac{m \Omega_\mathrm{rf} \omega_s}{e\sqrt{2}},
\end{equation}
where $\omega_s$ is psuedo-potential confinement frequency and $e$ the usual electron charge.  Similarly, the ideal quadrupole trap ($\epsilon=0$), has the same expression for $A$ with $\omega_s$ being the smaller radial confinement frequency.  Hence, matrix elements have a typical scale of $\epsilon \Theta(J)$.  For the $^1S_0$-to-$^3D_2$ clock transition in $^{176}$Lu$^+$, the magic rf at which micromotion shifts vanish is $\Omega_\mathrm{rf}\sim 2\pi\times 33\,\mathrm{MHz}$ \cite{arnold2018blackbody}, and the calculated value for $\Theta({^3}D_2)$ is $-1.77ea_0^2$ \cite{porsev2018clock}.  Using $\omega_s=2\pi\times 1\,\mathrm{MHz}$ then gives $\epsilon \Theta(J)/\hbar \sim 2\pi\times \,2\mathrm{kHz}$.  Lighter atoms often use higher $\Omega_\mathrm{rf}$ and/or have larger values of $\omega_s$.  Consequently, the value quoted for $^{176}$Lu$^+$ is reasonably indicative for other systems.  The exceptions would be those levels having an anomalously small quadrupole moment such as the ${^1}D_2$ level of Lu$^+$ \cite{porsev2018clock} or the $^2F_{7/5}$ level of Yb$^+$ \cite{huntemann2012high}.
\section{Effects within a single hyperfine level}
\subsection{Sideband modulation}
The time-varying frequency shift of a level will give rise to a sideband signal, as is the case for micromotion \cite{berkeland1998minimization} and the $z$ component of an ac magnetic field \cite{meir2018experimental,gan2018oscillating}.  The modulation index associated with the sideband is simply determined by the amplitude of the quadrupole shift divided by the trap drive frequency.  For $A=0$, the quadrupole-induced modulation index for a state $\ket{F,m_F}$ is
\begin{align}
\beta_Q&=\frac{\bra{F,m_F}H_Q\ket{F,m_F}}{\hbar \Omega_\mathrm{rf}}\\
&=\frac{m \omega_s}{\hbar e\sqrt{2}} C^{(2)}_{F,m_F} \Theta(J)\sin^2\beta\cos 2\alpha,
\end{align}
where $C^{(2)}_{F,m_F}$ is
\begin{multline}
C^{(2)}_{F,m_F}=(-1)^{2F+I+J+m_F}(2F+1)\\
\times \begin{Bmatrix}F & F & 2\\J & J &I\end{Bmatrix}\begin{pmatrix}F & 2 & F\\ m_F & 0 & -m_F \end{pmatrix} \begin{pmatrix}J & 2 & J\\ -J & 0 & J \end{pmatrix}^{-1}.
\end{multline}
Typically, $\Theta(J)\sim e a_0^2$ and $C^{(2)}_{F,m_F}\lesssim 1$.  Consequently $\beta_Q\lesssim 10^{-4}$ in most circumstances of interest and typically less than the modulation index arising from thermally-induced intrinsic micromotion \cite{keller2015precise}. 
\subsection{Resonant Zeeman coupling}
\label{RZC}
The quadrupole field can induce a precession of the spin when the Zeeman splitting between levels is resonant with the trap drive rf.  Since the quadrupole field can drive both $\Delta m=\pm1$, and $\pm2$ transitions, resonances occur when the Zeeman splitting between neighbouring $m$-states matches either $\Omega_\mathrm{rf}$ or $0.5\Omega_\mathrm{rf}$. Of interest are the $\Delta m=\pm 1$ transitions as this resonance can be used to measure trap-induced ac-magnetic fields \cite{gan2018oscillating}.  Accurate assessment of the magnetic field from the measured coupling strength would have to take into account the contribution from the quadrupole field.

To illustrate, consider the $F=5$ level of $^3D_2$ in $^{176}$Lu$^+$.  This level has the largest $g_F$ factor among all available clock levels and hence the smallest static field required to obtain resonance.  Within this level, the transition from $\ket{5,0}$ to $ \ket{5,\pm1}$ has the strongest magnetic coupling with a sensitivity of $21\,\mathrm{kHz/\mu T}$.  It also has the weakest quadrupole coupling within the $F=5$ manifold.  For $A=0$, the quadrupole coupling is given by
\begin{align}
\Omega_Q&=\frac{\bra{5,\pm1}H_Q\ket{5,0}}{\hbar}\nonumber\\
&=\sqrt{\frac{2}{3}}\frac{m \Omega_\mathrm{rf} \omega_s}{\hbar e\sqrt{2}}\left(\frac{\sqrt{5}}{26}\right)\Theta(J)\nonumber\\
&\qquad\times(\cos\beta\sin\beta \cos2\alpha\pm i \sin\beta \sin2\alpha).
\end{align}
Using $\omega_s=2\pi\times 1\,\mathrm{MHz}$,  $\Omega_\mathrm{rf}= 2\pi\times 33\,\mathrm{MHz}$, $\Theta({^3}D_2)=-1.77ea_0^2$ and neglecting the spatial orientation factor gives $\Omega_Q=2\pi\times 140\,\mathrm{Hz}$.  Going to a slightly higher static field would enable the $F=5$ level of $^1D_2$ level to be used instead.  This level has a $\sim20\%$ smaller sensitivity to ac magnetic fields, but the quadrupole moment is a factor 80 smaller \cite{porsev2018clock}.

In principle, the resonance at $0.5\Omega_\mathrm{rf}$ allows the quadrupole moment to be measured as first pointed out by Itano \cite{ItanoQuadExp}.  As this resonance only involves $|\Delta m|=2$ transitions, it cannot be driven by an ac magnetic field and only depends on the quadrupole coupling.  From trap frequency measurements, $A$ and $\epsilon$ can be accurately measured and maximising the coupling strength as a function of magnetic field direction fixes the orientation factor.  Hence the maximised coupling strength could then be related directly to the quadrupole moment.  A proof-of-principle demonstration using $^{138}$Ba$^+$ is given in section~\ref{experiment}.
\subsection{Off-resonant coupling}
When the Zeeman splittings do not match $\Omega_\mathrm{rf}$, off-resonant coupling modifies the Zeeman splitting.  In the case of an ac-magnetic field, the shift of each state is proportional to $m$ and can be viewed as a modified $g$-factor for the hyperfine level of interest.  In analogy with the ac Stark shift, the shift of level $m$ is given by
\begin{multline}
\frac{\delta E}{\hbar} =
-\sum_{\Delta m} \Bigg(\frac{|\bra{F,m+\Delta m}H_Q\ket{F,m}|^2}{2\hbar^2}\\
\times\frac{\omega_z \Delta m}{(\omega_z \Delta m)^2-\Omega_\mathrm{rf}^2}\Bigg),
\end{multline}
where $\omega_z=g_F \mu_B B_0/\hbar$ is the Zeeman splitting between neighbouring $m$-states.  The expression gives rise to terms proportional to $m$ and $m^3$.  In the limit that the Zeeman splitting is much less than $\Omega_\mathrm{rf}$, the effect has a scale of $(\epsilon\Theta(J)/\Omega_\mathrm{rf})^2$, which is likely well below $10^{-8}$ in most circumstances.
\section{Clock shifts}
For clocks with a hyperfine structure, it is prudent to estimate the shift that the oscillating quadrupole will have on the clock frequency.  In the limit that the trap drive frequency is much smaller than the hyperfine splittings, the shift of an $m=0$ clock state is given by
\begin{equation}
\label{clockshift}
h\delta \nu_F =
-\sum_{\Delta m,F'\neq F} \frac{|\bra{F',\Delta m}H_Q\ket{F,0}|^2}{2(E_{F'}-E_F)}.
\end{equation}
The summation excludes $F'=F$ as contributions within a hyperfine level cancel for $m=0$ states.  Each value of $|\Delta m|$ in the summation has a unique orientation dependence determined by the appropriate term in Eqs.~\ref{Wigner}.  In general, the weighting between different values of $|\Delta m|$ are dependent on the hyperfine structure resulting in a rather complicated orientation dependence of the shift.   Moreover, the shift does not cancel with various averaging methods \cite{ItanoQuad, dube2005electric, barrett2015NJP}.

Weightings for each $|\Delta m|$ can be readily calculated and, in the case of Lu$^+$, the result of hyperfine averaging \cite{barrett2015NJP} is easily included.  Hyperfine averaging cancels contributions from $\Delta m=0$ terms, which maybe readily verified from Eq.~\ref{clockshift}.  The fractional frequency shift can then be written in the form
\begin{equation}
\label{ShiftEq}
\frac{\delta \nu}{\nu}= a(f_2(\alpha,\beta)+\eta f_1(\alpha,\beta)),
\end{equation}
where $f_k(\alpha,\beta)$ is the magnitude squared of the appropriate orientation dependence for $|\Delta m|=k$ taken from Eqs.~\ref{Wigner}.  For $A=0$ this would be Eq.~\ref{linear1} and Eq.~\ref{linear2} for $f_1(\alpha,\beta)$ and $f_2(\alpha,\beta)$ respectively.

From quadrupole moments calculated in \cite{porsev2018clock} and measured hyperfine splittings \cite{kaewuam2017laser,kaewuam2018laser}, values of $a$ and $\eta$ can be readily calculated for a given trap setup. In table~\ref{LuShifts} values are tabulated using $A=0$, $\omega_s=2\pi\times 1\,\mathrm{MHz}$, and $\Omega_\mathrm{rf}= 2\pi\times 33\,\mathrm{MHz}$.  For all three clock transitions $\eta\approx-0.2$ giving maximum and minimum values of 1 and $\eta$, respectively, for $f_2(\alpha,\beta)+\eta f_1(\alpha,\beta)$.  
\begin{table}
 \caption{Parameters for Eq.~\ref{ShiftEq} determining the fractional frequency shifts of clock transitions in $^{176}$Lu$^+$.  Values are calculated using $\omega_s=2\pi\times 1\,\mathrm{MHz}$, $\Omega_\mathrm{rf}= 2\pi\times 33\,\mathrm{MHz}$ and quadrupole moments calculated in \cite{porsev2018clock}.}
 \label{LuShifts}
\begin{ruledtabular}
\begin{tabular}{c c c}
 \hspace{0.5cm}Transition & $a\;(10^{-19})$ & $\eta$ \hspace{0.5cm}\\
 \hline
\hspace{0.5cm}$^1S_0\leftrightarrow{^3}D_1$ & 1.28 & \hspace{0.25cm}-0.199 \hspace{0.5cm} \\
\hspace{0.5cm}$^1S_0\leftrightarrow{^3}D_2$& -0.90 & \hspace{0.25cm}-0.197 \hspace{0.5cm} \\
\hspace{0.5cm}$^1S_0\leftrightarrow{^1}D_2$ & 2.34[-4] & \hspace{0.25cm}-0.212 \hspace{0.5cm} \\
 \end{tabular}
 \end{ruledtabular}
 \end{table}

In a linear ion chain it would be desirable or indeed necessary to have $\beta\approx \cos^{-1}(1/\sqrt{3})$ to cancel quadrupole shifts induced by neighbouring ions.  In this case $f_2(\alpha,\beta)+\eta f_1(\alpha,\beta)\approx (3+\cos(4\alpha))/10$.  Consequently, the trap-induced rf quadrupole shift will be below $10^{-19}$ in any realistic circumstances.

Although the shifts are not likely to be a limitation in any foreseeable future, they could be assessed using resonant Zeeman coupling discussed in sect.~\ref{RZC}. Dominant contributions to the shift arise from the $|\Delta m|=2$ couplings, which have the same scaling factors and orientation dependence as the coupling strength of the resonance that occurs when the Zeeman splitting matches $0.5\Omega_\mathrm{rf}$.  Measurement of this resonant coupling strength could then provide a reasonable estimate of the associated clock shifts.
\section{Measuring the quadrupole coupling}
\label{experiment}
Coupling from the ac quadrupole field confining the ion can be observed when the Zeeman splitting of a level supporting a quadrupole moment matches $0.5\Omega_\mathrm{rf}$.  In this section a proof-of-principle demonstration is given using the $S_{1/2}-D_{5/2}$ clock transition at 1762\,nm in $^{138}$Ba$^+$.  Theoretical estimates of the $D_{5/2}$ quadrupole moment have been estimated by a number of researchers \cite{itano2006quadrupole, sahoo2006relativistic, sur2006electric, jiang2008electric} and the $g$-factor of $\sim 6/5$ allows the resonant condition to be meet at easily achievable fields ($\lesssim 1\,\mathrm{mT}$).  

The experiment is performed in a linear Paul trap similar to that used for previous work \cite{kaewuam2017laser,arnold2018blackbody}. The trap consists of two axial endcaps separated by $\sim2\,\mathrm{mm}$ and four rods arranged on a square with sides $1.2\,\mathrm{mm}$ in length. All electrodes are made from $0.45\,\mathrm{mm}$ electropolished copper-beryllium rods. Radial confinement is provided by a $20.585\,\mathrm{MHz}$ radio-frequency (rf) potential applied to a pair of diagonally opposing electrodes via a helical quarter-wave resonator.  With only differential voltages on the dc electrodes to compensate micromotion, the measured trap frequencies of a single $^{138}$Ba$^+$ are $(\omega_x,\omega_y,\omega_z)\sim 2\pi\times(990,895,112)\,\mathrm{kHz}$, with the trap axis along $z$.  Measurements described here were carried out in this configuration.  Ideally, with only an rf confinement, $\omega_z=\omega_x-\omega_y$, as can be readily verified from Eq.~\ref{principalEP}.  Deviations from this indicate an asymmetry resulting in field curvatures from the micromotion compensation potentials.  

To maximise the coupling to the quadrupole field, a magnetic field is aligned along the trap axis such that the principle and laboratory axes are aligned, that is, $\beta\approx0$ and $\alpha$ can be arbitrarily set to zero.  In this configuration the coupling strength for $\Delta m=\pm2$ transitions depends only on $\epsilon$.  Since the dc confinement from the micromotion compensation would not significantly affect the radial confinement, $\epsilon$ can be estimated by Eq.~\ref{epsilon} with $\omega_s$ given by the mean of $\omega_x$ and $\omega_y$. Moreover, for $\beta \ll 1$, the coupling strength varies quadratically with $\beta$ and is therefore insensitive to the exact alignment of the field to the trap axis.  However, limited optical access prevents optical pumping to a particular $m_J$ ground state, which limits population transfer to 0.5 when driving the clock transition and diminishes signal-to-noise.

\begin{figure}
\begin{center}
  \includegraphics[width=0.25\textwidth]{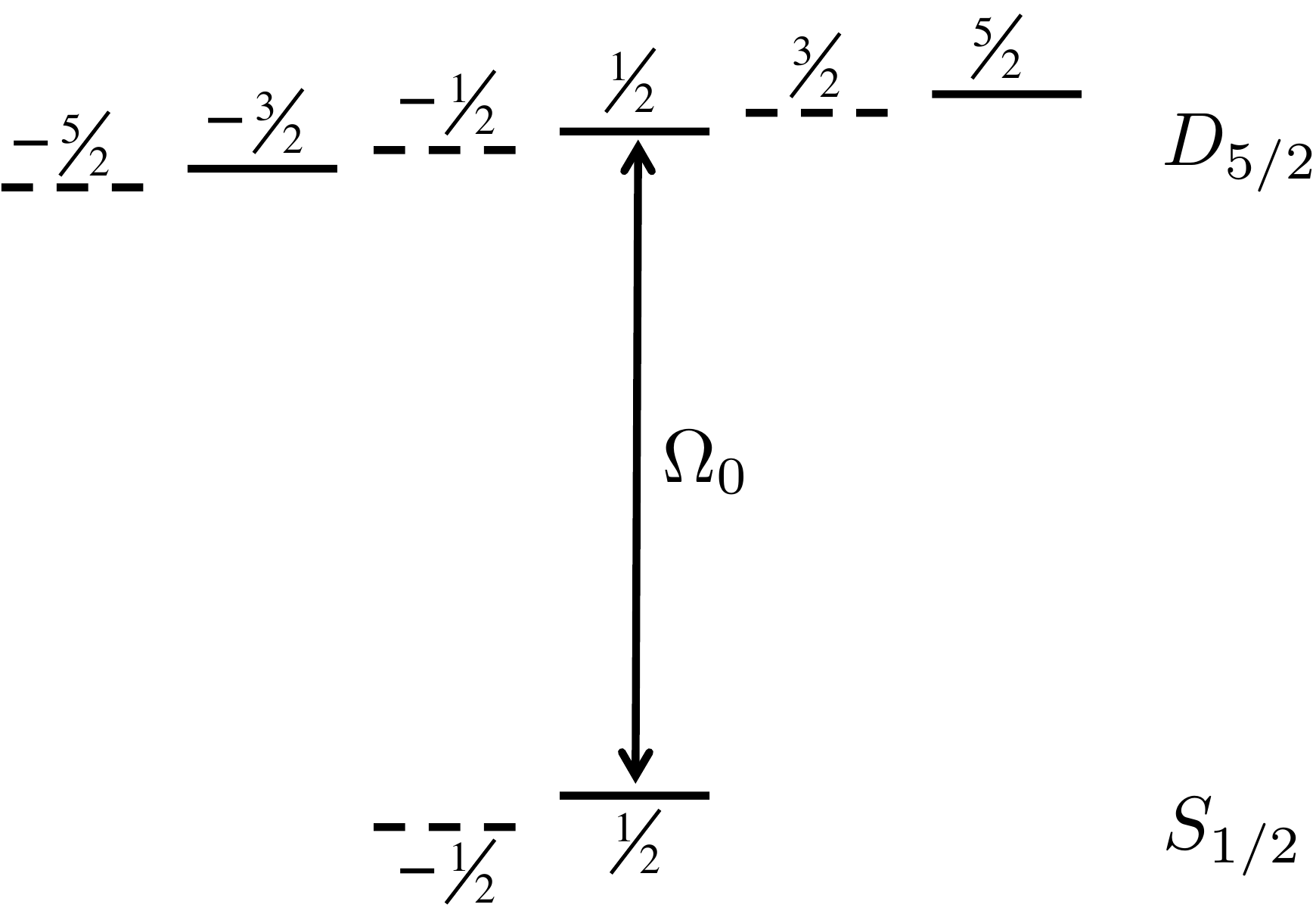}
  \caption{Level structure of $^{138}$Ba$^+$ used in the measurement of the quadrupole moment of $D_{5/2}$.  The clock laser drives $\ket{S_{1/2},m=1/2}$ to $\ket{D_{5/2},m=1/2}$ with coupling strength $\Omega_0$.  The oscillating quadrupole field confining the ion couples $\ket{D_{5/2},1/2}$ to $\ket{D_{5/2},-3/2}$ and $\ket{D_{5/2},5/2}$. Dashed levels do not contribute to the dynamics.}
  \label{BaFig}
\end{center}
\end{figure}

As depicted in Fig.~\ref{BaFig}, we consider driving the optical transition from $\ket{S_{1/2},m=1/2}$ to $\ket{D_{5/2},m=1/2}$.  The excited state Zeeman splitting $\omega_z=g_D\mu_B B_0/\hbar$ is set to one half the trap drive frequency $\Omega_\mathrm{rf}$, resulting in a resonant quadrupole coupling to $\ket{D_{5/2},5/2}$ and  $\ket{D_{5/2},-3/2}$.  With $\beta=0$, the Hamiltonian can be written
\begin{equation}
\label{Eq:Hamiltonian}
H/\hbar = \begin{pmatrix} -\Delta & \frac{\omega_\mathrm{Q}}{\sqrt{10}} & 0 & 0\\  \frac{\omega_\mathrm{Q}}{\sqrt{10}}  & 0 &  \frac{3 \omega_\mathrm{Q}}{5\sqrt{2}} &  \frac{1}{2}\Omega_{0}\\ 0 &   \frac{3 \omega_\mathrm{Q}}{5\sqrt{2}} & \Delta &  0\\  0 &\frac{1}{2}\Omega_{0} &0& \delta \end{pmatrix},
\end{equation}
where $\omega_Q=\epsilon \Theta/\hbar$ is the characteristic strength of the quadrupole coupling, $\Omega_0$ is the coupling strength of the clock laser, and the rotating wave approximation has been used for both the quadrupole and laser coupling.  This assumes the detunings $\Delta=\Omega_\mathrm{rf}-2\omega_z$, and $\delta=\omega-\left(\omega_0+(\omega_z-\omega'_z)/2\right)$ are both small with respect to $\Omega_\mathrm{rf}$ and the Zeeman-shifted clock frequency $\omega_0+(\omega_z-\omega'_z)/2$, respectively. In this last expression $\omega'_z=g_S\mu_B B_0/\hbar$ is the Zeeman splitting between the $S_{1/2}$ states.

\begin{figure}
\begin{center}
  \includegraphics[width=\columnwidth]{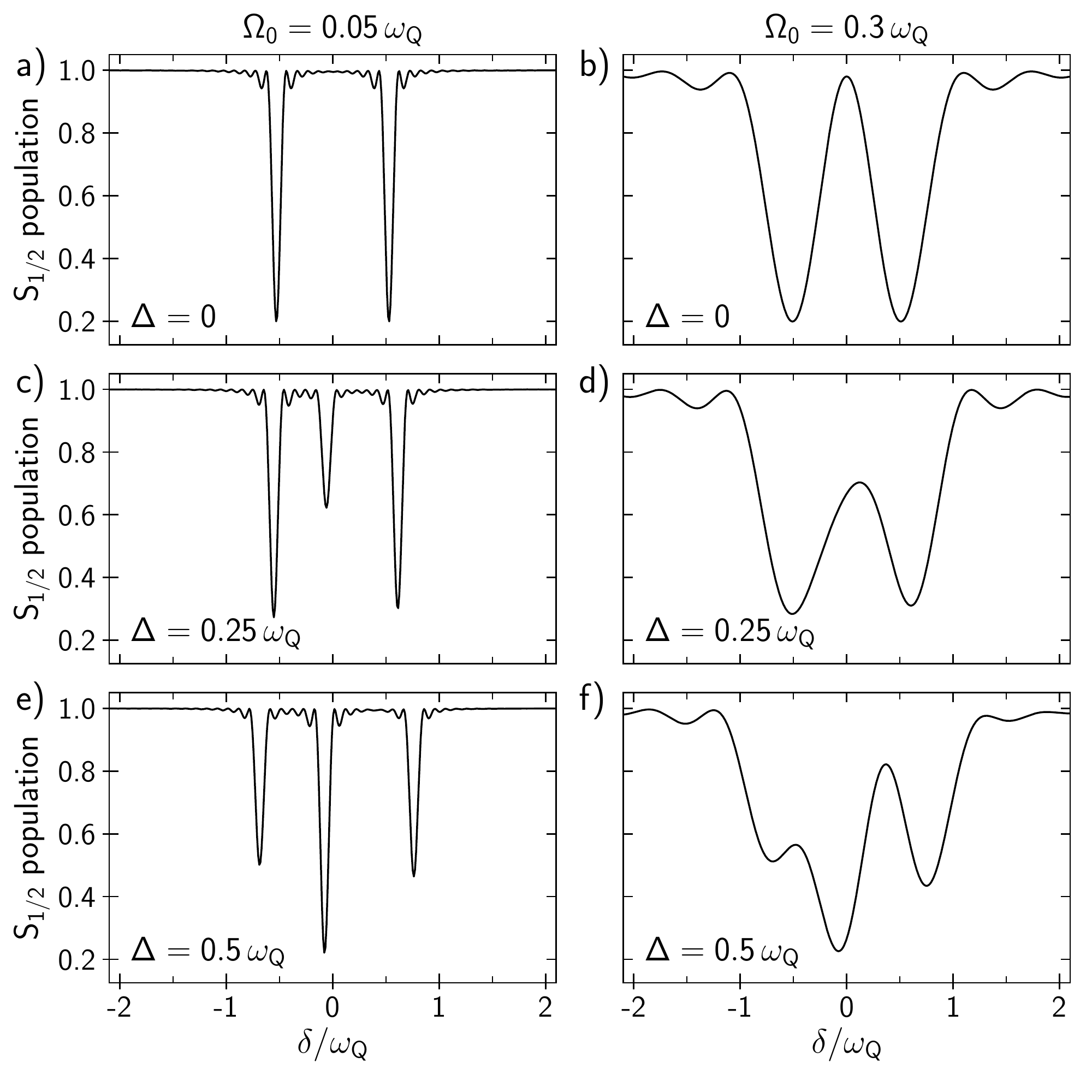}
  \caption{Theoretical curves derived from Eq.~\ref{Eq:Hamiltonian} for different values of $\Omega_0$ and $\Delta$.  Plots on the left and right have $\Omega=0.05\omega_Q$ and $0.3\omega_Q$ respectively.  Detunings $\Delta$ from top to bottom are $0,0.25\omega_Q,$ and $0.5\omega_Q $. All plots use an interrogation time $\tau=\pi/\Omega_0$, which is the $\pi$-time when the Zeeman splitting is far from the quadrupole resonance.}
  \label{AutlerTownes}
\end{center}
\end{figure}

When $\Omega_0\ll \omega_Q$, the quadrupole coupling results in an Autler-Townes triplet \cite{autler1955stark} as illustrated in Fig.~\ref{AutlerTownes}.  Plots are given for $\Omega_0=0.05\,\omega_Q$ (left) and $\Omega_0=0.3\,\omega_Q$ (right) and for detunings $\Delta=0, 0.25\,\omega_Q,$ and $0.5\,\omega_Q$.  In all cases the clock interrogation time is given by $\tau=\pi/\Omega_0$.  Note that there are only two lines when $\Delta=0$ due to the fact that one of the dressed states has no amplitude in $\ket{D_{5/2},1/2}$ as is evident from Eq.~\ref{Eq:Hamiltonian} for this case.

\begin{figure*}
\begin{center}
  \includegraphics[width=\linewidth]{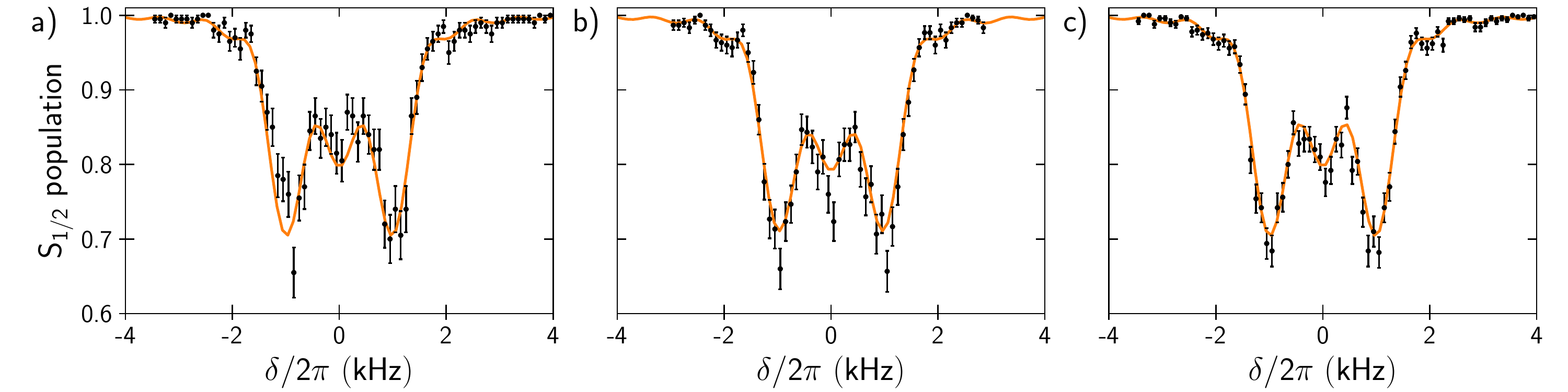}
  \caption{Experimentally observed signals as a function of the clock laser detuning from the  $\ket{S_{1/2},1/2}$ to $\ket{D_{5/2},1/2}$ transition.  Each data point represents 200, 300, and 500 experiments for plots (a), (b), and (c) respectively.  Solid curves are fits using the rms spread of an assumed Gaussian magnetic field noise, and quadrupole coupling strength, $\omega_Q$, as fitting parameters.  Fit parameters are tabulated in Table~\ref{dataFits}.}
  \label{quadExp}
\end{center}
\end{figure*}

Experimentally, magnetic field noise limits the clock probe time.  This limits the resolution of the Autler-Townes splitting, as illustrated in Fig.~\ref{AutlerTownes}, and the experimentally observed signal is further degraded by the changing detunings induced by magnetic field variations as seen in three data runs shown in Fig.~\ref{quadExp}.  For all data sets $\Delta \approx 0$ and the clock probe time is set to $1.2\,\mathrm{ms}$, which is the $\pi$-time for the transition when the Zeeman splitting is far from the quadrupole resonance.  Each data point represents 200, 300, and 500 experiments for plots (a), (b), and (c) respectively.   The solid curve given in each plot is derived using a Gaussian distributed magnetic field noise, which is assumed constant within a single experiment.  The standard deviation ($\sigma$) of magnetic field deviations about the mean and $\omega_Q$ are determined by a $\chi^2$-fit to the data.  Fit parameters and the reduced-$\chi^2$ for each dataset are given in Table~\ref{dataFits}.  The quoted errors in the fit parameters are determined by standard methods using the covariance matrix from the fit and have used error scaling to allow for the larger values of $\chi^2$.  Offset detunings have not been included in the fits as their inclusion does not significantly change the $\chi^2$ or the fitted values of $\sigma$ and $\omega_Q$. 
\begin{table}
 \label{dataFits}
\caption{Fit parameters and reduced-$\chi^2$ for the curves given in Fig.~\ref{quadExp}.  Errors are determined by standard methods using the covariance matrix from the fit and have used error scaling to allow for the larger values of $\chi^2$.}
\begin{ruledtabular}
\begin{tabular}{l c c c c}
\hspace{0.25cm} Plot & N & $\omega_Q/2\pi$ (Hz) & $\sigma$ (nT) & $\chi_\nu^2$ \hspace{0.25cm} \\
 \hline
\hspace{0.25cm} Fig.~\ref{quadExp}(a) & 200 & $1708\,(24)$ & $18.2\,(1.1)$  & 1.15 \hspace{0.25cm} \\
\hspace{0.25cm} Fig.~\ref{quadExp}(b) & 300 & $1662\,(19)$ & $18.4\,(0.8)$ & 1.11 \hspace{0.25cm} \\
\hspace{0.25cm} Fig.~\ref{quadExp}(c) & 500 & $1713\,(16)$ & $18.2\,(0.7)$ & 1.48 \hspace{0.25cm} \\
 \end{tabular}
 \end{ruledtabular}
 \end{table}
The reduced $\chi^2$ of 1.49 for plot (c) indicates a statistically poor fit and the difference in $\omega_Q$ for plot (b) is statistically significant.  This is likely due to slow variations of the average magnetic field over the timescales of a full data scan, which are not captured by the model.  The fitted values of $\sigma$ are consistent with the coherence times and stabilities observed when servoing on the clock transition.  However, this does not account for a possible linear drift of the mean magnetic field over the duration of the scan, which would increase or decrease the fitted value of $\omega_Q$ depending on sign of the drift.  Based on clock servo data taken over 6 hours, this drift is unlikely to be more than $20\,\mathrm{nT}$ over the duration of any scan shown in Fig.~\ref{quadExp}. This corresponds to an error of approximately 1.4\% or $24\,\mathrm{Hz}$ in $\omega_Q$.  Adding this error in quadrature with the largest error from the fits and taking the mean of the fitted values gives $\omega_Q=2\pi\times 1.694(35)\,\mathrm{kHz}$ as an estimate of the coupling strength.  

To estimate the quadrupole moment $\Theta$, an estimate of $\epsilon$ and hence $\omega_s$ is needed.  As noted, $\omega_s$ is ideally given by the mean of $\omega_x$ and $\omega_y$ and $\omega_z=\omega_x-\omega_y$.  Taking the discrepancy between measured values of $\omega_z$ and $\omega_x-\omega_y$ as a conservative error estimate gives $\omega_s=2\pi\times943(17)\,\mathrm{kHz}$.  Using Eq.~\ref{epsilon} and the definition of $\omega_Q$ then gives $\Theta=3.229(89) e a_0^2$.  A comparison with available theoretical estimates is given in Table~\ref{quadComparison}.  Although the estimated value is in fair agreement with the theoretical value of $\Theta=3.319\,e a_0^2$ given in \cite{jiang2008electric}, a more precise experimental value would be desirable to test the accuracy of the theory.
\begin{table}
 \label{quadComparison}
\caption{Comparison of the current experimental and theoretical estimates of the electric quadrupole moment for the $5d\,D_{5/2}$ level of Ba$^+$.  All values of given in units of $e a_0^2$.}
\begin{ruledtabular}
\begin{tabular}{l c c c c}
\hspace{0.25cm} Present & Ref.~\cite{itano2006quadrupole} & Ref.~\cite{sahoo2006relativistic} & Ref.~\cite{sur2006electric} & Ref.~\cite{jiang2008electric} \hspace{0.25cm} \\
 \hline
\hspace{0.25cm} 3.229(89) & 3.379 & 3.42(4) & 3.382(61)  & 3.319(15) \hspace{0.25cm}
 \end{tabular}
 \end{ruledtabular}
 \end{table}
 
The implementation here was is limited by magnetic field noise and stability of the mean magnetic field over longer timescales.  Modest improvements in magnetic field noise would allow individual dressed states to be resolved.  With consideration of both $\ket{S_{1/2},\pm1/2}$ to $\ket{D_{5/2},1/2}$ transitions, the clock laser could be servoed to the outer dressed states associated with each transition.  This would allow (i) the ground-state splitting to be servoed to the correct value as determined by known $g$-factors \cite{marx1998precise,knab1993experimental,hoffman2013radio} and $\Omega_\mathrm{rf}$, (ii) the clock laser to be maintained on line center, and (iii) the dressed-state splitting to be measured to a precision limited by the integration time.  Determination of the quadrupole moment would then be limited by the determination of the rf confinement potential.  This typically dominates over the dc contribution and could be assessed much more accurately than done here.   Measurement at the 0.1\% uncertainty should be achievable using this approach.

\section{Summary}
In this paper, the effects of an oscillating quadrupole field on atomic energy levels have been considered.  General expressions for the interaction matrix elements have been given and can be used to calculate effects for any given set up.  This work generalises the results given in \cite{ItanoQuad} for the first order quadrupole shift from a static field.  It also complements the recent discussion on ac magnetic field effects driven by the same trapping fields considered here \cite{gan2018oscillating}.  Although the quadrupole effects are small and not likely to limit clock performance, their possible influence on the assessment of ac magnetic fields should be considered if the levels involved support a quadrupole moment.  

A proof-of-principle measurement of the quadrupole moment using the resonant coupling induced by the rf confinement has also been demonstrated.  Without any special magnetic field control an accuracy of $\sim3\%$ has been achieved.   Modest improvement in magnetic field control and a rigorous assessment of the trap rf confinement would significantly improve accuracy.  Other methods have utilised entangled states within decoherence free subspaces \cite{roos2006designer} or dynamic decoupling \cite{shaniv2016atomic}, both of which have achieved inaccuracies at the $\sim0.5\%$ level.  In any case one is limited by the size of the interaction and the ability to characterize the potential.  The method here utilizes the dominant coupling produced by the rf confinement, which is also less sensitive to stray fields.  Additionally, the method is technically easy to implement. 

\begin{acknowledgements}
We thank Wayne Itano for suggesting the possibility of using the resonance condition $2\omega_z=\Omega_\mathrm{rf}$ to measure the quadrupole moment and bringing our attention to his earlier work. This work is supported by the National Research Foundation, Prime Ministers Office, Singapore and the Ministry of Education, Singapore under the Research Centres of Excellence programme. This work is also supported by A*STAR SERC 2015 Public Sector Research Funding (PSF) Grant (SERC Project No: 1521200080). T. R. Tan acknowledges support from the Lee Kuan Yew post-doctoral fellowship.
\end{acknowledgements}
\bibliography{QuadrupoleFields}
\bibliographystyle{unsrt}
\end{document}